# Electronic phase diagram of iron chalcogenide superconductors FeSe$_{1-x}$S$_x$ and FeSe$_{1-y}$Te$_y$

Shaobo Liu[1,2], Jie Yuan[1,5,6], Soonsang Huh[3,4], Hanyoung Ryu[3,4], Mingwei Ma[1,6], Wei Hu[1,2], Dong Li[1,2], Sheng Ma[1,2], Shunli Ni[1,2], Peipei Shen[1,2], Kui Jin[1,2,5,6], Li Yu[1,2,6*], Changyoung Kim[3,4], Fang Zhou[1,2,6*], Xiaoli Dong[1,2,5,6*], Zhongxian Zhao[1,2,5,6]

[1] *Beijing National Laboratory for Condensed Matter Physics and Institute of Physics, Chinese Academy of Sciences, Beijing 100190, China*

[2] *School of Physical Sciences, University of Chinese Academy of Sciences, Beijing 100049, China*

[3] *Center for Correlated Electron Systems, Institute for Basic Science, Seoul 08826, Republic of Korea*

[4] *Department of Physics and Astronomy, Seoul National University, Seoul 08826, Republic of Korea*

[5] *Key Laboratory for Vacuum Physics, University of Chinese Academy of Sciences, Beijing 100049, China*

[6] *Songshan Lake Materials Laboratory, Dongguan, Guangdong 523808, China*

\* Correspondence to: fzhou@iphy.ac.cn (F.Z.); li.yu@iphy.ac.cn (L.Y.); dong@iphy.ac.cn (X.L.D.)

## Abstract

Here we establish a combined electronic phase diagram of isoelectronic FeSe$_{1-x}$S$_x$ (0.19 ≥ $x$ ≥ 0.0) and FeSe$_{1-y}$Te$_y$ (0.04 ≤ $y$ ≤ 1.0) single crystals. The FeSe$_{1-y}$Te$_y$ crystals with $y$ = 0.04 – 0.30 are grown by a hydrothermal ion-deintercalation (HID) method. Based on combined experiments of the specific heat, electrical transport, and angle-resolved photoemission spectroscopy, no signature of the tetragonal-symmetry-broken transition to orthorhombic (nematic) phase is observed in the HID FeSe$_{1-y}$Te$_y$ samples, as compared with the FeSe$_{1-x}$S$_x$ samples showing this transition at $T_s$. A ubiquitous dip-like temperature dependence of the Hall coefficient is observed around a characteristic temperature $T^*$ in the tetragonal regimes, which is well above the superconducting transition. More importantly, we find that the superconducting transition temperature $T_c$ is positively correlated with the Hall-dip temperature $T^*$ across the FeSe$_{1-x}$S$_x$ and FeSe$_{1-y}$Te$_y$ systems, suggesting that the tetragonal background is a fundamental host for the superconductivity.

- ***Introduction.*** The superconductivity in iron chalcogenide Fe(Te,Se) emerges as the long-range antiferromagnetic (AFM) order in the parent compound FeTe is suppressed [1], similar to the superconductivity in iron pnictides [2,3]. The non-magnetic iron chalcogenide FeSe superconductor is unusual in that a nematic electronic order has been observed [4-9], which is associated with a rotational-symmetry-breaking transition from a tetragonal to an orthorhombic phase at $T_s \sim 90$ K. The superconductivity in FeSe coexists with the nematic order, and shows a rather low transition temperature $T_c$ (~9 K). Nevertheless, the superconductivity can be strongly enhanced by high pressure ($T_c \sim 38$ K), chemical intercalation ($T_c \sim 30 - 50$ K), or fabrication of monolayer FeSe ($T_c \sim 65$ K). The nematicity is completely suppressed in these cases, and significant doping effects can also be involved. Tuning the superconducting (SC) and electronic properties by the isovalent substitutions of chalcogen elements S and Te for Se in FeSe is a pristine way to probe the origin of superconductivity, with the doping effect reduced to minimum. The resultant $FeSe_{1-x}S_x$ [10] and $FeSe_{1-y}Te_y$ [11-13] series consist only of the superconducting/conducting blocks each comprising one iron plane sandwiched between two chalcogen planes, in absence of the interlayers of guest atoms or molecules. Therefore, these simple iron-based compounds provide a unique platform to investigate the intrinsic ingredients for the unconventional superconductivity in a wide isoelectronic phase region. Particularly, it is highly desirable to obtain experimental evidence for the link between the superconductivity and normal-state electronic property by tracking their respective dependences on the isovalent substitution.

In this study, we succeed in synthesizing a series of tellurium-substituted $FeSe_{1-y}Te_y$ single crystals, covering the substitution range of $0.04 \leq y \leq 0.30$, by our recently developed hydrothermal ion-deintercalation (HID) method [14]. Single crystals of sulfur-substituted $FeSe_{1-x}S_x$ with $x = 0$ (FeSe), 0.07 and 0.13, and $FeSe_{1-y}Te_y$ with higher $y = 0.61$, 0.89 and 1.0, are also obtained by chemical-vapor-transport (CVT) method [15,16] and self-flux growth [17] with post annealing [18], respectively. Further, we establish a combined phase diagram of $FeSe_{1-y}Te_y$ ($0.04 \leq y \leq 1.0$) and $FeSe_{1-x}S_x$ ($0.19 \geq x \geq 0.0$) systems based on systematic characterizations. Contrasting with the presence of the orthorhombic/nematic phase below $T_s$ in stoichiometric FeSe and substituted FeSe$_{1-x}$S$_x$, no signature of the electronic nematicity is detectable in the HID FeSe$_{1-y}$Te$_y$ samples from the specific heat and electrical transport measurements. The absence of nematicity is further supported by the angle-resolved photoemission spectroscopy (ARPES) measurements, with no $d_{xz}/d_{yz}$ band splitting observed around the Brillouin zone (BZ) corners. In addition, the band structure around the BZ center is significantly altered for tetragonal HID FeSe$_{1-y}$Te$_y$, as compared with the previous reports for tetragonal FeSe$_{1-y}$Te$_y$ with higher Te contents ($y > 0.5$). Our observations support that the highly mobile electron state emerging in the nematic regime does not contribute to the superconductivity. Interestingly, a ubiquitous dip-like temperature dependence of the Hall coefficient is observed around a characteristic temperature $T^*$ in the tetragonal regimes, which is well above $T_c$. More importantly, we find that the superconducting transition temperature $T_c$ is closely coupled to the Hall-dip temperature $T^*$ across the FeSe$_{1-x}$S$_x$ and FeSe$_{1-y}$Te$_y$ systems. Therefore, our results suggest that the underlying physics setting in at higher temperature in tetragonal background is fundamental for the origin of unconventional superconductivity in the multiband iron-based compounds.

**- Sample description.** FeSe$_{1-y}$Te$_y$ single crystals with $y = 0.04, 0.06, 0.11, 0.16, 0.22$ and $0.30$ were obtained through the soft-chemical HID process [14] (see also Supplementary Fig. S1). The as-grown HID FeSe$_{1-y}$Te$_y$ crystals are large in size (up to 14 mm; Supplementary Fig. S1c) and free from the chemical phase separation as present in the polycrystalline samples [12]. This allows us to readily identify the chemical phase and measure the intrinsic physical properties. The chemical compositions of all the samples were determined by inductively coupled plasma-atomic emission spectrometry. A monotonic, continuous crystal-lattice contraction from FeSe$_{1-y}$Te$_y$ to FeSe$_{1-x}$S$_x$ with decreasing $y$ or increasing $x$ was confirmed by x-ray diffraction (XRD) measurements. This is in agreement with the decrease of the ionic radius from Te, Se to S, and indicates that a continuous isovalent substitution is achieved. Correspondingly, a monotonic increase in the chemical compressive stress is expected from FeSe$_{1-x}$Te$_x$ to FeSe$_{1-y}$S$_y$. All our single crystals exhibit a single preferred (001) orientation. The superconducting transition was characterized by both the magnetic susceptibility ($\chi$) and in-plane resistivity ($\rho$) measurements, with the $T_c$ value determined from the onset temperature of the superconducting shielding signal in $\chi(T)$. The SC transition is sharp, with the width $\Delta T_c \leq 1$ K for FeSe$_{1-y}$Te$_y$ samples (except for a $\Delta T_c \sim 1.7$ K of $y = 0.30$ with the lowest $T_c$), and $\Delta T_c \leq 0.5$ K for FeSe$_{1-x}$S$_x$ samples. All the single crystals display a 100 % or large (~80 % for $y = 0.30$) superconducting shielding signal, demonstrating their bulk superconductivity. All the details can be found in Supplementary Figs. S3-5.

**- Nematicity and ARPES results.** The S-substituted FeSe$_{1-x}$S$_x$ undergoes a tetragonal to orthorhombic/nematic phase transition upon cooling to $T_s$. The value of $T_s$ is reduced with increasing substitution $x$, from maximal $T_s \sim 90$ K at $x = 0$ (FeSe) to ~35 K at $x = 0.16$, and finally vanishes at $x \gtrsim 0.17$ (refs. 19-21). The orthorhombic/nematic transition has been shown to manifest itself in many experiments, including a jump or kink at $T_s$ in the $T$-dependent specific heat (Fig. 1b) or resistivity (Fig. 2d-f), respectively, and a drop in the Hall coefficient at $T'$ just below $T_s$ (Fig. 2j-l). Recently, the nematic transition has been reported in CVT-grown FeSe$_{1-y}$Te$_y$ ($y \lesssim 0.4$) single crystals from the resistivity kink [22,23] and Hall-coefficient drop [22], where $T_s$ decreases with increasing $y$. However, in our HID-grown FeSe$_{1-y}$Te$_y$ ($0.04 \leq y \leq 0.30$) single crystals, none of the specific-heat jump (Fig. 1a), resistivity kink (Fig. 3a-c), and Hall-coefficient drop (Fig. 3g-i) characteristic of the transition is observed. The presence of the tetragonal phase down to the lowest measuring temperature has also been reported in other iron chalcogenide superconductors, such as binary charge-doped Fe$_{1-x}$Se (ref. 24) and molecule-intercalated high-$T_c$ (~42 K) (Li$_{1-x}$Fe$_x$OH)FeSe (ref. 25).

Then, we present the low-$T$ (6 K) electronic structures measured by ARPES for the representative HID FeSe$_{1-y}$Te$_y$ at $y = 0.22$ and CVT FeSe$_{1-x}$S$_x$ at $x = 0.13$, shown in Fig. 3. As can be seen from Fig. 4, the tetragonal FeSe$_{0.78}$Te$_{0.22}$ is intermediate between the strongly nematic FeSe ($x$ or $y = 0$) and optimal-$T_c$ tetragonal FeSe$_{1-y}$Te$_y$ ($y = 0.5$), while the optimal-$T_c$ nematic FeSe$_{0.87}$S$_{0.13}$ is away from FeSe in the other side at a comparable substitution interval. The Fermi-surface mapping is shown in Fig. 3a for FeSe$_{0.87}$S$_{0.13}$ and 3b for FeSe$_{0.78}$Te$_{0.22}$, where dashed blue/red lines mark the $\Gamma$-$M_1$ high-symmetry cuts. The intensity plots of FeSe$_{0.87}$S$_{0.13}$ around the $\Gamma$- and $M_1$-point are presented in Fig. 3c and 3e, respectively. The observed hole- and electron-like bands are guided by dashed curves in Fig. 3d (MDC second derivative; MDC: momentum distribution curve) and 3f (EDC second derivative; EDC: energy distribution curve). The Fermi level ($E_F$) crossings of the hole ($\alpha$) and electron ($\delta$) bands are demonstrated by the MDCs at $E_F$, the blue curves in the upper parts of Fig. 3c and 3e, respectively. The band structures around both $\Gamma$- and $M_1$-point for FeSe$_{0.87}$S$_{0.13}$ are consistent with previous ARPES results [4-9]. In accordance with the previous orbital-character assignments, the hole-bands $\alpha/\beta$ near $\Gamma$ and $\alpha'/\beta'$ near $M_1$ (ref. 9) are dominated by Fe $d_{xz}/d_{yz}$ orbital characters, the hole-band $\gamma$ near $\Gamma$ by the $d_{xy}$ orbital character. The corresponding spectra of FeSe$_{0.78}$Te$_{0.22}$ are shown in Fig. 3g/h ($\Gamma$-cut) and 3i/j ($M_1$-cut), with the bands marked by solid curves in Fig. 3h/j. The hole-band $\varepsilon$ and electron-band $\kappa$ cross the Fermi level near $\Gamma$ and $M_1$, respectively, as demonstrated by the MDCs at $E_F$ (red curves in the upper parts of Fig. 3g and 3i). The electron-band $\kappa$ is more clearly visible from the $M_2$-cut data (Supplementary Fig. S6c). The splitting of $d_{xz}/d_{yz}$ bands around the $M_1$-point has been widely used to identify the orbital-nematic order [4-9], as shown in Fig. 3e/f/k for the present nematic FeSe$_{0.87}$S$_{0.13}$. The corresponding splitting energy $\Delta_M$ can be estimated from the EDC at the $M_1$-point (blue curve in Fig. 3k). By comparison, the band structure around $M_1$ for the tetragonal HID FeSe$_{0.78}$Te$_{0.22}$ (Fig. 3i/j/k) is similar to that [26,27] reported previously for the tetragonal ($y \geq 0.5$; ref. 28) flux FeSe$_{1-y}$Te$_y$ samples. Therefore, the ARPES results support the specific heat and electrical transport characterizations for the presence (FeSe$_{0.87}$S$_{0.13}$) or absence (FeSe$_{0.78}$Te$_{0.22}$) of the nematicity.

Previous ARPES studies [26,27,29] of tetragonal FeSe$_{1-y}$Te$_y$ with $y > 0.5$ have revealed three hole-bands around $\Gamma$ in the vicinity of $E_F$. In the present tetragonal FeSe$_{0.78}$Te$_{0.22}$, we identify only two of them with the $d_{xz}/d_{yz}$ orbital characters, marked as $\varepsilon/\eta$ here; a third hole-band ($d_{xy}$) is not discernible. The $\varepsilon$ ($d_{xz}$) and $\eta$ ($d_{yz}$) hole-bands are resolved by the MDC second derivative around $\Gamma$-cut (Fig. 3h) and the overall EDC second derivative along $\Gamma$-$M_1$ direction (Fig. 3l), respectively. We find that, compared to the previous ARPES results, the $\eta$ ($d_{yz}$) hole-band does not cross the Fermi level, such that the two hole-bands $\eta$ ($d_{yz}$) and $\varepsilon$ ($d_{xz}$) cross at ~50 meV below $E_F$ (Fig. 3h/l). These characteristics of band structure around $\Gamma$ differ from the previous observations of the highly substituted FeSe$_{1-y}$Te$_y$ (refs. 26,27,29). Therefore, our results indicate that reducing the Te content in tetragonal FeSe$_{1-y}$Te$_y$ mainly affects the band structure around the $\Gamma$ point, while that around the $M$ points is not significantly altered.

*- Hall and resistivity results.* Now we switch to an overall view of the in-plane transport data measured for the FeSe$_{1-x}$S$_x$ and FeSe$_{1-y}$Te$_y$ single crystals. The temperature-dependent Hall coefficient $R_H(T)$ of both HID FeSe$_{1-y}$Te$_y$ (Fig. 2g-i) and CVT FeSe$_{1-x}$S$_x$ (Fig. 2j-l) shows a dip around a characteristic temperature $T^*$ in their tetragonal regimes, featuring an obvious upturn in $R_H(T)$ below $T^*$. $R_H(T)$ of tetragonal HID FeSe$_{1-y}$Te$_y$ keeps on rising with cooling below the Hall-dip $T^*$ (Fig. 2g-i). In contrast, $R_H(T)$ of nematic FeSe$_{1-x}$S$_x$ exhibits a complex behavior consistent with previous observations [30-32]: The rising trend below $T^*$ collapses to a dramatic drop starting at $T'$ (Fig. 2j-l). Here we emphasize that the Hall-drop $T'$ is slightly below the structural $T_s$ (Fig. 2j-l; Fig. 4). In highly substituted tetragonal FeSe$_{1-y}$Te$_y$ samples, the Hall coefficient generally displays persistent positive values [33] (see also Supplementary Fig. S9), with weakly negative $R_H(T)$ values observed only in a very limited region at low $T$ (refs. 33,34). In their tetragonal regimes, all the FeSe$_{1-x}$S$_x$ and FeSe$_{1-y}$Te$_y$ samples exhibit a linear Hall resistivity $\rho_{xy}(H)$ (Supplementary Figs. S8-9), consistent with the previous results and the nearly compensated metal character [30,31,35,36]. In the nematic regime of FeSe$_{1-x}$S$_x$, a nonlinearity in $\rho_{xy}(H)$ develops below $T'$ (Supplementary Fig. S8), accompanying the drop in Hall coefficient (Fig. 2j-l). This deviation from the linear $\rho_{xy}(H)$ has been attributed to the emergence of an additional small and highly mobile electron pocket associated with the nematicity [30,35].

The in-plane resistivity $\rho(T)$ of FeSe$_{1-x}$S$_x$ samples shows a metallic behavior over the whole measuring temperature range (Fig. 2d-f), similar to previous reports [30,32,37-39]. By comparison, a broad hump appears in $\rho(T)$ of HID FeSe$_{1-y}$Te$_y$ samples (Fig. 2a-c), and the hump position seems to shift to lower temperature with increasing $y$. This crossover to non-metallic behavior with heating is similar to the crossover observed previously by ARPES at $y = 0.56$, ascribed to a strong orbital-dependent band renormalization [27]. Here a noticeable non-metallic upturn also develops in the low-$T$ resistivity of the HID samples at higher $y$ levels (see Fig. 2c for $y = 0.22$). The extrapolated residual resistivity for the nematic FeSe$_{1-x}$S$_x$ samples is at least one order of magnitude smaller than that for the tetragonal FeSe$_{1-y}$Te$_y$ samples at $y \geq 0.5$. For example, the residual resistivity is only 0.016 mΩ·cm for the present optimal-$T_c$ nematic FeSe$_{1-x}$S$_x$ ($x = 0.13$), while that for the previous optimal-$T_c$ tetragonal FeSe$_{1-y}$Te$_y$ ($y = 0.5$) [34] is about 0.16 mΩ·cm. Therefore the impurity/defect scatterings are much weaker in the nematic FeSe$_{1-x}$S$_x$ samples.

*- Phase diagram and discussion.* In Fig. 4, we plot the temperature vs. substitution ($x$, $y$) phase diagram of the FeSe$_{1-x}$S$_x$ ($0.19 \geq x \geq 0.0$) and FeSe$_{1-y}$Te$_y$ ($0.04 \leq y \leq 1.0$) systems, by summarizing the present characterizations and some related data from literature. The normal-state region is superimposed with a contour plot of the Hall coefficient. This joint phase diagram clearly shows a double-dome-like evolution of $T_c$, maximized at $x = 0.13$ ($T_c = 10.8$ K) for FeSe$_{1-x}$S$_x$ and at $y \sim 0.5$ ($T_c = 14.4$ K; ref. 40) for FeSe$_{1-y}$Te$_y$. Similar superconducting dome of FeSe$_{1-x}$S$_x$ has also been reported in earlier work [37].

In the nematic regime ($0.17 \gtrsim x \geq 0$), the electronic nematicity of FeSe$_{1-x}$S$_x$ is enhanced with decreasing $x$ (ref. 7). Both the structural $T_s$ and Hall-drop $T'$ rise correspondingly, reaching their maximal values at stoichiometric FeSe. This trend is also followed by a temperature scale slightly below $T'$ (the white/zero contour line), at which the Hall coefficient changes sign from positive to negative with cooling for $0.1 \gtrsim x \geq 0$. Moreover, a higher $T'$ corresponds to a more negative $R_H(T)$ value at low $T$ (see also Fig. 2j-l). These observations support that the emergence of the high-mobility electron carriers below the Hall-drop temperature $T'$ is inherently connected with the nematic ordering [30,35]. In the superconducting state, as the system is tuned to cross the boundary ($x \sim 0.17$) into the nematic regime, $T_c$ of FeSe$_{1-x}$S$_x$ rises initially with decreasing $x$ then starts to decline at $x = 0.13$, compared to the continuous enhancement of the nematicity. As crossing from the strongly nematic FeSe at $x = 0$ into the tetragonal HID FeSe$_{1-y}$Te$_y$ region, $T_c$ continues to decline smoothly with increasing $y$. That is, the changing rate of $T_c$ with substitution remains the same as crossing the FeSe$_{1-x}$S$_x$/FeSe$_{1-y}$Te$_y$ boundary, despite the sudden disappearance of the high-mobility electron state (as characterized by the concurrently vanishing structural $T_s$ and Hall-drop $T'$ upon entering the tetragonal region). These observations suggest that the highly mobile electron channel associated with the nematicity does not contribute to the superconducting channel, similar to the recent result of charge-doped $Fe_{1-x}Se$ system [24].

For most compensated tetragonal $FeSe_{1-y}Te_y$ samples, the Hall coefficient is positive at low $T$ near $T_c$, indicating the hole dominance. Particularly for tetragonal HID samples showing no Hall-drop $T'$, the value of Hall coefficient becomes strongly positive towards $T_c$. We find that $T_c$ of HID $FeSe_{1-y}Te_y$ is increased with the enhancement of the positive $R_H(T)$ near $T_c$ (see Fig. 2g-i). This highlights the hole component relevance to the superconductivity. In nematic $FeSe_{1-x}S_x$, however, such a strong hole dominance near $T_c$ otherwise seen by the Hall measurement turns out to be either significantly weakened (the light red contour near $x = 0.13$), or completely blanketed (the blue contour for $0.13 > x \geq 0$), by the emergent high-mobility electrons below Hall-drop $T'$. Since the highly mobile electron state developing at the lower temperature in the nematic regime does not contribute to the superconductivity, the respective roles in the superconductivity played by the nearly compensated holes and electrons, already existing at higher temperatures in the tetragonal regimes, both deserve in-depth investigation. Additionally, in tetragonal high-$T_c$ $(Li_{1-x}Fe_xOH)FeSe$ system with the compensated character manifesting as a persistent electron dominance [25,41,42], it has also been shown that both the electrons and holes could contribute to the superconductivity [42].

Furthermore, the phase diagram reveals a close resemblance between the double-dome shape of $T_c$ and the non-monotonic evolution of $T^*$ in the tetragonal regimes, across the $FeSe_{1-x}S_x$ and $FeSe_{1-y}Te_y$ systems. Namely, the superconducting transition temperature $T_c$ is pushed up as the Hall-dip temperature $T^*$ is increased. It is noteworthy that a similar $T_c$ scaling with $T^*$ has also been observed in a series of single-crystalline films [42] and single crystal [41] of high-$T_c$ $(Li_{1-x}Fe_xOH)FeSe$. In Fig. 5, we show the positive correlations between $T_c$ and $T^*$ found in the three iron chalcogenide systems. At higher $T^*$, the $(Li_{1-x}Fe_xOH)FeSe$ intercalates display much higher $T_c/T^*$ ratios than the isoelectronic samples, likely due to the enhanced extra doping effect from the interlayers as the charge reservoir [25]. Nevertheless, as $T^*$ is reduced, the extrapolated $T_c$ values of all the three systems approach zero as well.

Therefore, this ubiquitous close link between the superconducting $T_c$ and Hall-dip $T^*$ implies that some generic ingredients essential for superconductivity have developed around $T^*$ in the tetragonal regimes, well above $T_c$. It is also important to note here that, in our earlier study of the highly two-dimensional high-$T_c$ (Li$_{1-x}$Fe$_x$OH)FeSe single crystal, we have observed a nearly linear magnetic susceptibility (extracted for the iron planes) and a linear in-plane resistivity, pointing to the presence of magnetic fluctuations [41,43] as further evidenced by later inelastic neutron scattering experiments [44,45]. Interestingly, both the linear magnetic susceptibility and resistivity start to appear below the Hall-dip temperature $T^*$. Here in proximity to the FeTe end, the superconductivity emerges as the AFM order below $T_N$ is suppressed [1]. The origin of iron-based superconductivity remains an issue under debate among the scenarios proposed so far, with the proposal of the spin-fluctuation-mediated pairing mechanism as one of them. Previous neutron studies have observed in the tetragonal regimes of both FeSe (at $T > T_s$) (ref. 46) and FeSe$_{1-y}$Te$_y$ (refs. 1,47) the magnetic fluctuations similar to that in iron pnictide superconductors. Further experimental and theoretical studies are certainly needed to pin down the underlying physics setting in below $T^*$ and the connection with the subsequent superconductivity at $T_c$.

Therefore, our results suggest the tetragonal background to be a fundamental host for the superconductivity. This leads to a straightforward understanding of a partial suppression of superconductivity in the nematic regime. It is obvious that the present nematic FeSe$_{1-x}$S$_x$ ($x$ = 0.13) shows an optimal $T_c$ (10.8 K) lower than that ($T_c$ = 14.4 K) of the previous optimal tetragonal FeSe$_{1-y}$Te$_y$ ($y$ = 0.5); in fact, at a given temperature scale of $T^*$ shared by the isoelectronic samples from the two systems, the nematic FeSe$_{1-x}$S$_x$ samples (blue-shaded area in Fig. 5) consistently exhibit a lower $T_c$ than the tetragonal FeSe$_{1-y}$Te$_y$ samples, despite having the same Hall-dip $T^*$ as, and the much weaker impurity/defect scatterings than, the tetragonal samples. Such a partial suppression of superconductivity is consistent with the recognized irrelevance to superconductivity of the emergent highly mobile electron state, and the reported orbital-selective suppression of possible pairing channels [9,48], in the nematic state. Moreover, a twofold-symmetric SC gap has been identified in the nematic FeSe (refs. 49-51) and FeSe$_{1-x}$S$_x$ (ref. 52), as compared with a modulated fourfold [53,54] or isotropic [55,56] SC gap symmetry observed in the tetragonal FeSe$_{1-y}$Te$_y$.

- *Conclusion.* We establish a combined electronic phase diagram of the isoelectronic iron chalcogenides from FeSe$_{1-x}$S$_x$ (0.19 ≥ $x$ ≥ 0) to FeSe$_{1-y}$Te$_y$ (0.04 ≤ $y$ ≤ 1.0). No signature of the electronic nematicity is observed in the HID-grown FeSe$_{1-y}$Te$_y$ (0.04 ≤ $y$ ≤ 0.30) single crystals. We show that reducing the Te content in tetragonal FeSe$_{1-y}$Te$_y$ significantly affects the band structure only around the $\Gamma$ point. Our observations support that the highly mobile electron state emerging at lower temperature in the nematic regime does not contribute to the superconductivity. We also show that there exists a ubiquitous dip-like temperature dependence of the Hall coefficient around a higher temperature scale $T^*$ in the tetragonal regimes, from the lower-$T_c$ FeSe$_{1-x}$S$_x$-FeSe$_{1-y}$Te$_y$ systems studied here to a series of iron-chalcogenide intercalates of high-$T_c$ (Li$_{1-x}$Fe$_x$OH)FeSe system as reported previously. More importantly, we find that the superconducting transition temperature $T_c$ is pushed up as the Hall-dip temperature $T^*$ is increased in these variants of iron-based superconductors. This ubiquitous close link between $T_c$ and $T^*$ implies that some generic ingredients essential for the superconductivity have developed well above $T_c$.

**Methods**
The XRD measurements were carried out at room temperature on an 18 kW MXP18A-HF diffractometer with Cu-K$_\alpha$ radiation, with a 2$\theta$ scanning step of 0.01 ° for the single crystals or 0.02 ° for the ground powders. The magnetic susceptibility data were measured on a Quantum Design MPMS-XL1 system with a tiny remnant field (< 4 mOe). The in-plane electrical transport measurements were performed on a Quantum Design PPMS-9 system with a six-probe configuration. The Hall signals were determined by subtracting the data in negative fields from that in positive fields with mathematical interpolation. The heat capacity data were measured on the PPMS-9 system by a thermal relaxation method. The ARPES experiments were conducted on a house-made ARPES system (Instrument II) at Center for Correlated Electron System, Institute of Basic Science (Seoul National University, Republic of Korea). The setup was equipped with a SCIENTA DA30 electron analyzer and a Helium discharge lamp as light source (He-I, $h\nu$ = 21.218 eV). The total energy resolution was set to be better than 15 meV, and the angular resolution was about 0.5°. The single crystals were cleaved at 5 K and measured at 6 K in an ultrahigh vacuum with basal pressure better than 5×10$^{-11}$ mbar.

**Acknowledgements**

We would like to thank Drs. L. H. Yang, J. Li, and P. Zheng for technical assistance. This work was supported by the National Key Research and Development Program of China (Grant Nos. 2016YFA0300300 and 2017YFA0303003), the National Natural Science Foundation of China (Grant Nos. 11834016 and 11888101), and the Strategic Priority Research Program and Key Research Program of Frontier Sciences of the Chinese Academy of Sciences (Grant No. QYZDY-SSW-SLH001). The work at SNU was supported by the Institute for Basic Science in Korea (Grant No. IBS-R009-G2).

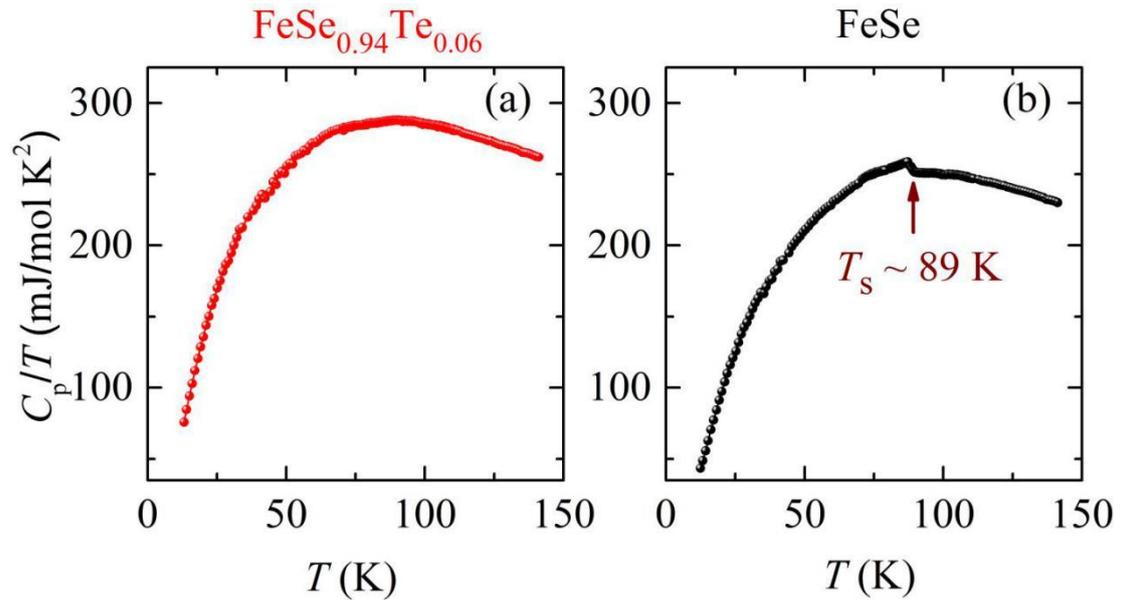

FIG. 1. Temperature dependences of the specific heat for the single crystals of (a) representative tetragonal HID $FeSe_{0.94}Te_{0.06}$ and (b) nematic CVT FeSe.

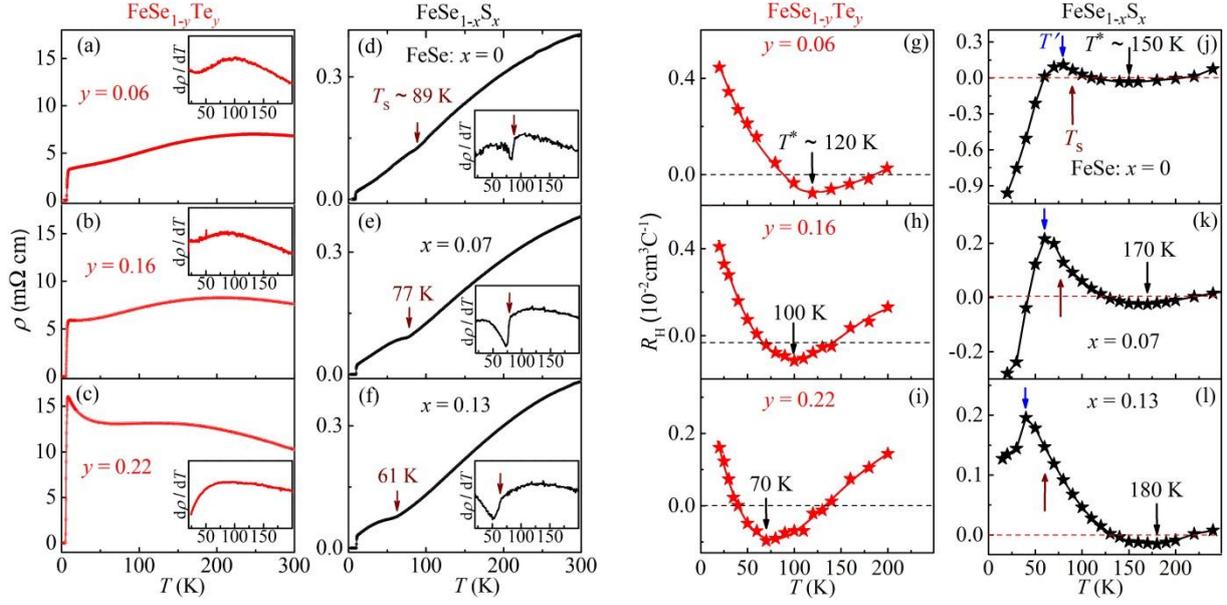

FIG. 2. (a-c) and (d-f) show the temperature-dependent in-plane resistivity for representative tetragonal HID FeSe$_{1-y}$Te$_y$ and nematic CVT FeSe$_{1-x}$S$_x$ single crystals, respectively. The data of FeSe$_{1-y}$Te$_y$ at other Te substitutions are given in Supplementary Fig. S7. Insets: The temperature derivatives of the resistivity exhibit clear drops at $T_s$ for the nematic FeSe$_{1-x}$S$_x$ samples. (g-i) and (j-l) are the temperature-dependenct Hall coefficients of these samples, obtained from the fit to the linear Hall resistivity $\rho_{xy}(H)$ or the field derivative of $\rho_{xy}(H)$ at the zero-field limit in the case of nonlinear $\rho_{xy}(H)$. All the $\rho_{xy}(H)$ data are presented in Supplementary Figs. S8-9.

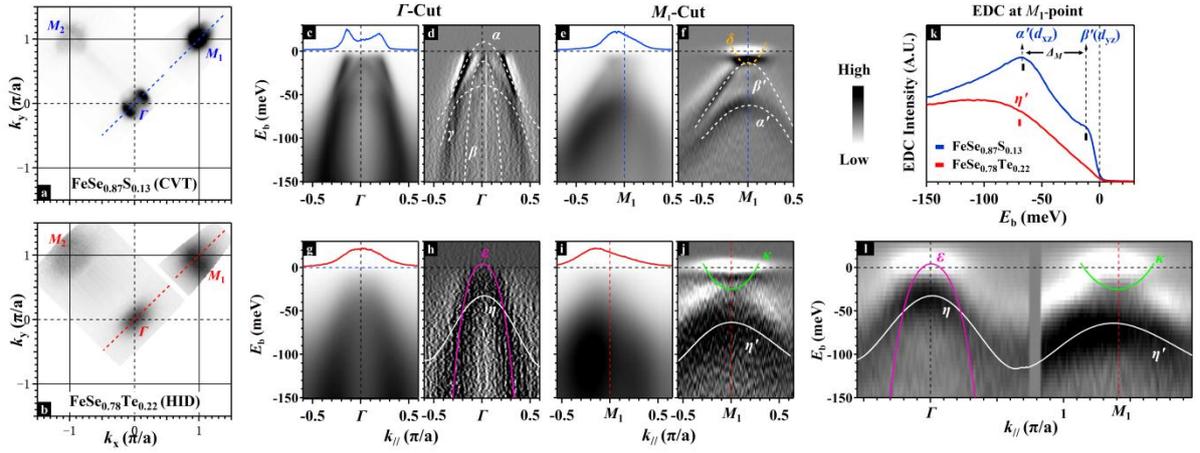

FIG. 3. (a) and (b) show the ARPES Fermi-surface mapping in the 2-Fe BZ for nematic CVT FeSe$_{1-x}$S$_x$ ($x$ = 0.13) and tetragonal HID FeSe$_{1-y}$Te$_y$ ($y$ = 0.22), respectively. The dashed blue/red lines indicate the $\Gamma$-$M_1$ high-symmetry direction. The intensity plots of FeSe$_{0.87}$S$_{0.13}$ are shown in (c) for the $\Gamma$-cut and (e) for the $M_1$-cut, with their MDC/EDC second derivatives presented in (d) and (f), respectively. In the upper parts of (c) and (e) are the MDCs at $E_F$. (g)-(j) show the corresponding data for FeSe$_{0.78}$Te$_{0.22}$. (k) The blue EDC at $M_1$ characterizes the band splitting energy $\Delta_M$ for nematic FeSe$_{0.87}$S$_{0.13}$. (l) Overall EDC second derivative along the $\Gamma$-$M_1$ direction for FeSe$_{0.78}$Te$_{0.22}$. All the ARPES data were measured at $T$ = 6 K with $h\nu$ = 21.218 eV from He-I discharging.

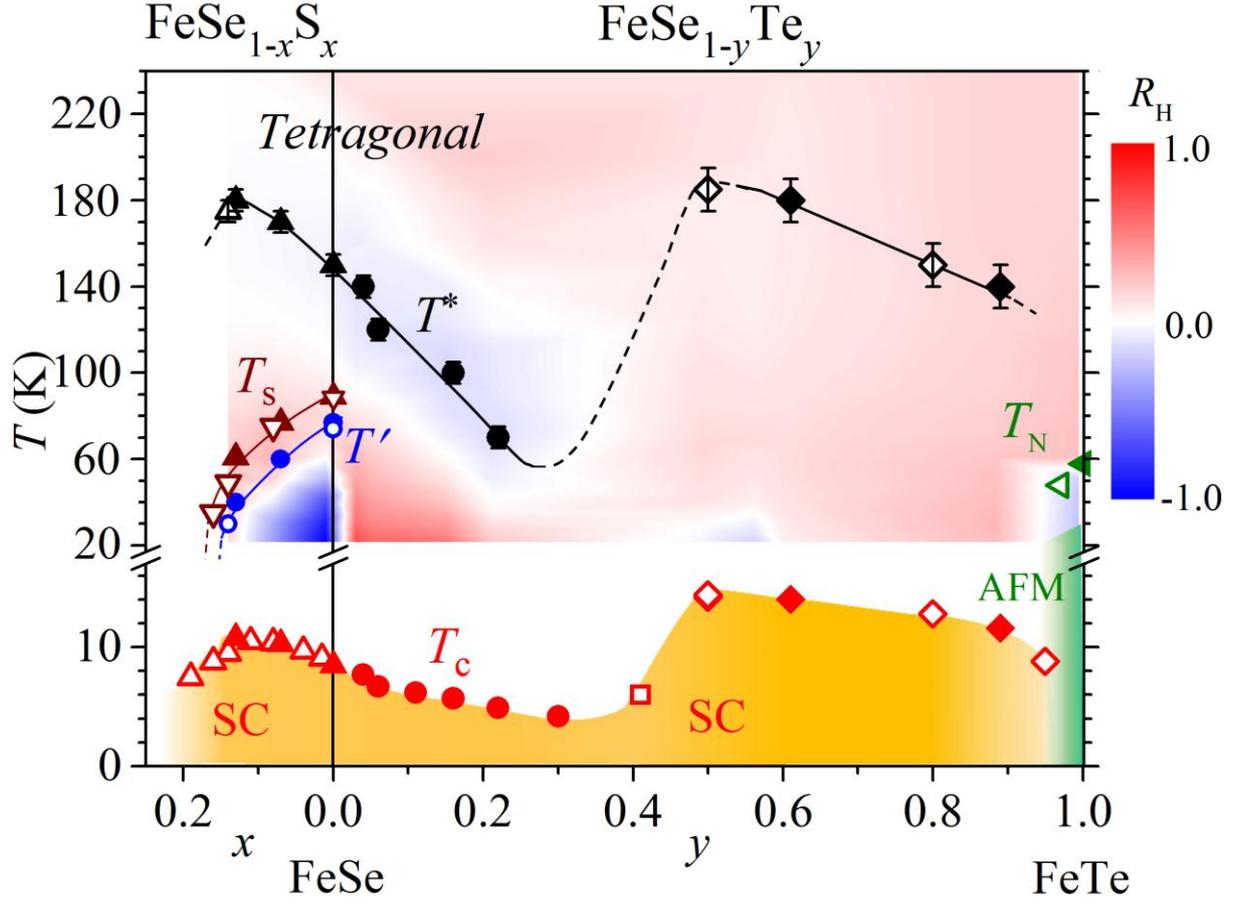

FIG. 4. $T$-$x/y$ phase diagram of the FeSe$_{1-x}$S$_x$ and FeSe$_{1-y}$Te$_y$ systems, superimposed by a contour plot of the Hall coefficient $R_H$ in the normal state. The present results are denoted by filled symbols, the data from literature by open symbols. The solid/dashed lines are drawn as guide to the eye. (FZ for the red open square is short for the floating-zone method.)

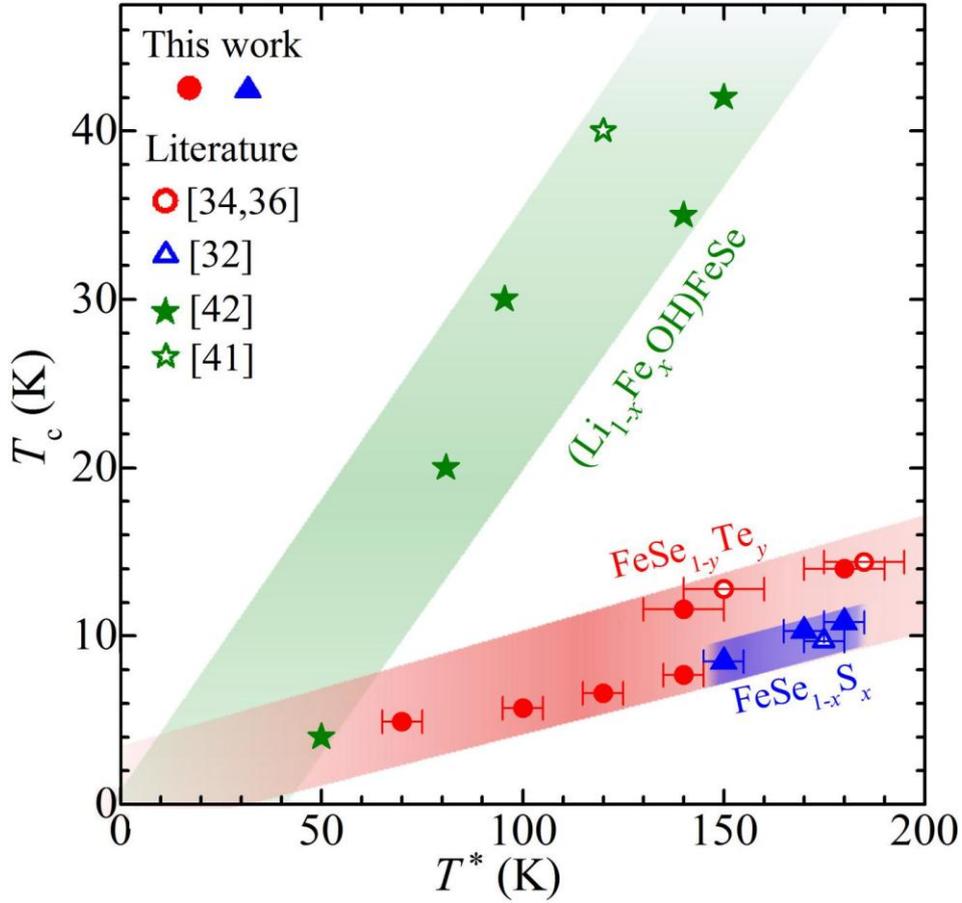

FIG. 5. Positive correlations between the superconducting $T_c$ and Hall-dip $T^*$ observed in the three iron-chalcogenide systems, with the colored shadings showing the trends. The filled/open circles and triangles represent the present/previous results of FeSe$_{1-y}$Te$_y$ and FeSe$_{1-x}$S$_x$ single crystals, respectively. The filled and open stars stand for the previous data of single-crystalline films [42] and single crystal [41] of high-$T_c$ (Li$_{1-x}$Fe$_x$OH)FeSe system, respectively. Note: The $T_c$ values of the films were taken as the zero-resistivity onset temperatures [42].